\documentstyle[twoside,fleqn,espcrc2,psfig]{article}
\title{Effects of Chemical Potential on Hadron Masses in the 
Phase Transition Region 
\thanks{Talk presented by T.Takaishi}}
\author{ {\large QCD-TARO Collaboration:}
 Ph.~de~Forcrand${}^a$,      M.~Garc{\'\i}a~P\'erez${}^b$,
 T.~Hashimoto${}^c$,      S.~Hioki${}^d$, Y.~Liu${}^e$
 H.~Matsufuru${}^{j}$,    
 O.~Miyamura${}^e$,
 A.~Nakamura${}^g$,          I.-O.~Stamatescu${}^{f,h}$,
 T. Takaishi${}^i$
 and  T.~Umeda${}^e$ \\ 
\vspace{4mm}
${}^a$ ETH-Z\"urich, CH-8092 Z\"urich, Switzerland  \\
${}^b$ Dept. F\'{\i}sica Te\'orica, Universidad Aut\'onoma de Madrid,
 E-28049 Madrid, Spain \\ 
${}^c$ Dept. of Appl. Phys., Fac. of Engineering,
        Fukui Univ., Fukui 910-8507, Japan  \\
${}^d$ Dept. of Physics, Tezukayama Univ.,
        Nara 631-8501, Japan  \\
${}^e$ Dept. of Physics, Hiroshima Univ.,
        Higashi-Hiroshima 739-8526, Japan  \\
${}^f$ Inst. Theor. Physik, Univ. of Heidelberg,
        D-69120 Heidelberg, Germany  \\
${}^g$ Res. Inst. for Inform. Sci. and Education, Hiroshima Univ.,
        Higashi-Hiroshima  739-8521, Japan  \\
${}^h$ FEST, Schmeilweg 5, D-69118 Heidelberg, Germany \\
${}^i$ Hiroshima University of Economics,
        Hiroshima 731-0192, Japan  \\ 
${}^j$ RCNP, Osaka University, Osaka 567, Japan\\}

\begin{document}
\begin{abstract}
We study the response of hadron masses with respect to chemical potential
at $\mu=0$.
Our preliminary results of the pion channel show that $\partial m/\partial \mu$ 
in the confinement phase 
is significantly larger than that in the deconfinement phase, which is 
consistent with the chiral restoration. 
\end{abstract}

\maketitle

\section{Introduction}

As suggested by QCD sum rule analysis \cite{HATSU}, hadron masses may be 
affected by density effects.  This may explain some results of 
heavy ion collision experiments such as dilepton spectra 
and J/$\Psi$ suppression.

It is difficult to introduce density effects in lattice QCD calculations
due to the well-known ``complex action" problem.
Here we calculate the response of hadron masses to chemical potential,
$\partial m/\partial \mu$,
on dynamical configurations with $\mu=0$.
Since simulations are done at $\mu=0$, there is no difficulty  in
obtaining $\partial m/\partial \mu$.
We investigate the dependence of $\partial m/\partial \mu$ with the temperature.

\section{Formulation}
We use  2 flavors of staggered quarks. 
The effective action to simulate $N_f$ fermion flavors
is
\begin{equation}
S_{eff}=S_G + S_F 
\end{equation}
where $S_G$ is the standard plaquette action and
\begin{equation}
S_F=\frac{N_f}{4}{\rm Tr} \ln M(U,\mu)
\end{equation} 
where $M(U,\mu)$ is the staggered fermion Matrix.

The zero momentum hadron correlation function $G(t)$ is given by
\begin{equation} 
G(t)=\sum_{x}<H(x,t)H(0,0)^\dagger>
\end{equation}
and
\begin{eqnarray}
&& <H(x,t)H(0,0)^\dagger> \\ \nonumber 
&& = \int dU H(x,t)H(0,0)^\dagger \exp(-S_{eff}) /Z  \\ \nonumber
\label{eq2}
\end{eqnarray}
where $Z$ is the partition function.

Taking a derivative of the hadronic correlator 
with respect to $\mu$.

\begin{eqnarray}
\label{eq3}
&&\!\!\!\frac{\partial <H(x,t)H(0,0)^\dagger> }{\partial \mu} \ = \ \ <\frac{\partial C(x,t)}{\partial \mu}> \\ 
&&\!\!\!-<C(x,t) \frac{\partial S_F}{\partial \mu}>  
+ <C(x,t)><\frac{\partial S_F}{\partial \mu}>. \nonumber
\end{eqnarray}
where $C(x,t)=H(x,t)H(0,0)^\dagger$.
We calculate eq. (\ref{eq3}) on dynamical configurations with $\mu=0$.
In the case of $\mu=0$ eq. (\ref{eq3}) can be simplified 
using the following facts:

(A) $\partial S_F/\partial \mu$ corresponds to the fermion number operator.
Thus, the average of the fermion number operator at $\mu=0$ is zero: 
$<\frac{\partial S_F}{\partial \mu}>=0$.

(B) On each configuration the value of $\partial S_F/\partial \mu$ 
is purely imaginary \cite{IMA}. 
Thus, the value of \mbox{$<C(x,t) \frac{\partial S_F}{\partial \mu}>$} is 
also purely imaginary  
provided that the operator $C(x,t)$ is real. 
This is indeed the case if we consider $C(x,t)$ for mesons made up of 
degenerate quarks.

Using the facts (A) and (B) above we derive
\begin{equation}
\frac{\partial <H(x,t)H(0,0)^\dagger>}{\partial \mu}=
<\frac{\partial C(x,t)}{\partial \mu}>
\end{equation}
for mesons consisting of degenerate quarks.

In the spectral representation, 
\begin{equation}
G(t)=\sum_i A_i \cosh(m_i(t-N_t/2)).
\label{eq4}
\end{equation}

Taking a derivative of eq. (\ref{eq4}) with respect to $\mu$ we obtain
\begin{eqnarray}
\label{eq5}
&&\frac{\partial G(t)}{\partial \mu} = 
\sum_i [\frac{\partial A_i}{\partial \mu} \cosh(m_i(t-N_t/2)) \\ \nonumber
&&+\frac{\partial m_i}{\partial \mu}A_i(t-N_t/2)
\sinh(m_i(t-N_t/2))].
\end{eqnarray}

Our procedure to obtain $\partial m/\partial \mu$ is as follows.
First we determine $A_i$ and $m_i$ by fitting correlation function 
data to eq. (\ref{eq4}). 
Substituting the values of $A_i$ and $m_i$ into eq. (\ref{eq5})
we fit the data of $\frac{\partial G(t)}{\partial \mu}$ to
eq. (\ref{eq5}). Then we obtain $\partial m_i/\partial \mu$ and 
$\partial A_i/\partial \mu$ as fitting parameters.

\section{Definition of $\partial /\partial \mu$}
We study the two flavor case ($u$ and $d$ quarks).
In this case, we have two independent chemical potentials,
$\mu_u$ and $\mu_d$.  Instead, the following combinations are convenient,
$\mu_S=(\mu_u+\mu_d)/2$   and $\mu_V=(\mu_u-\mu_d)/2$
whith $\mu_S$ the usual  chemical potential corresponding to
baryon number.  Then derivatives with respect to $\mu_S$ and $\mu_V$   
are
\begin{eqnarray}
{\partial \over \partial \mu_S} =
{\partial \over \partial \mu_u}+{\partial \over \partial \mu_d} =   
{\partial \over \partial \mu_u}-{\partial \over \partial \mu_{\bar{d}}}
\end{eqnarray}

\begin{eqnarray}
{\partial \over \partial \mu_V} =
{\partial \over \partial \mu_u}-{\partial \over \partial \mu_d} =
{\partial \over \partial \mu_u}+{\partial \over \partial \mu_{\bar{d}}}\ \ .
\end{eqnarray}

For degenerate systems of $u$ and $d$ quarks,
\begin{eqnarray}
{\partial C_{u\bar{d}} \over \partial \mu_S}=
{\partial C_{u\bar{d}} \over \partial \mu_u}-
{\partial C_{u\bar{d}} \over \partial \mu_{\bar{d}}} = 0 \ \ .
\end{eqnarray}
at $\mu_u=\mu_d=0$.
In this study we analize $\partial /\partial \mu_V$ which 
gives non-trivial results even with degenerate quarks.
In the following $\partial / \partial \mu$ stands for
$\partial /\partial \mu_V$.

\begin{figure}[ht]
\centerline{\psfig{figure=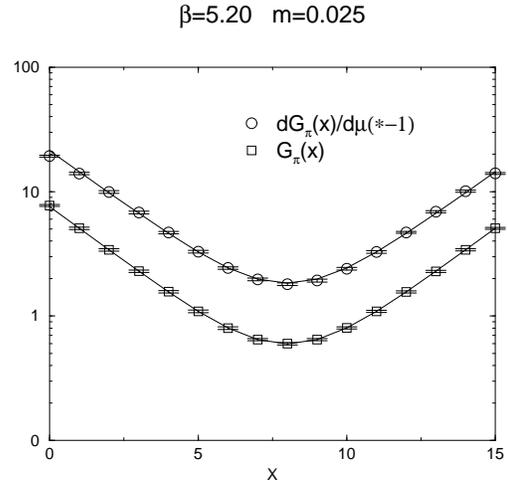,height=6.5cm}}
\vspace{-5mm}
\caption{The pion correlation function, $G_{\pi}(x)$ and its derivative 
with respect to the 
chemical potential, $\frac{\partial G_{\pi}(x)}{\partial \mu}$
at $\beta=5.20$.
$\frac{\partial G_{\pi}(x)}{\partial \mu}$ gives negative values.
To plot them in logarithmic scale, they are multiplied by -1.
Single pole fitting results are also shown, represented by solid lines.
}
\end{figure}

\section{Preliminary results}
We present preliminary results of $\partial m/\partial \mu$ for 
$N_f=2$  staggered quarks.
Simulations are done on a lattice of size $16\times8\times8\times4$ at $m_q=0.025$
with $\beta=5.20$, 5.26, 5.32 and 5.34.
We use the R-algorithm to generate configurations.
The finite temperature transition occurs at $\beta\approx 5.28$ \cite{FINITE}
and the above $\beta$ values are translated to $T/T_c=0.90,0.97,1.06$ 
and 1.09 respectively.

We measure the pion screening mass.
The quark propagator is calculated with $m_q=0.025$ (light) and 0.25 (heavy).
Then we construct the pion correlator with light-light and light-heavy quarks.

Fig. 1 shows the pion (light-light) correlation function $G_{\pi}(x)$ 
and its derivative with respect to  $\mu$ at $\beta=5.20$.
We perform single pole fit for the data, which turned out to be 
sufficient for the pion channel. 

\begin{figure}
\centerline{\psfig{figure=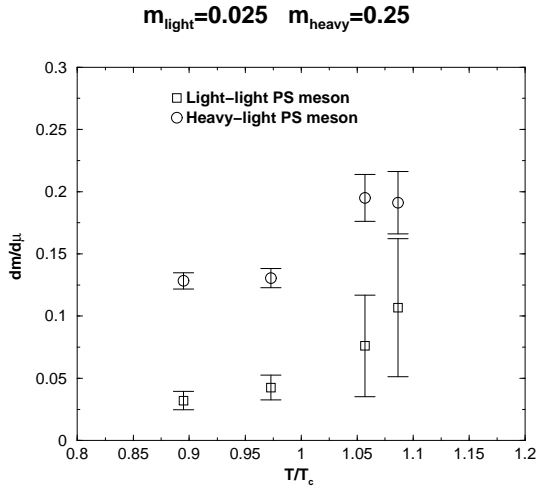,height=6.5cm}}
\vspace{-5mm}
\caption{$\partial m/\partial \mu$ of
light-light and heavy-light pseudoscalar mesons 
as a function of $T/T_c$.}
\end{figure}

Fig. 2 shows $\partial m/\partial \mu$ as a function of $T/T_c$.
Despite the large errors we observe a systematic tendency
towards raising the derivative of m above $T_c$.

\begin{figure}
\centerline{\psfig{figure=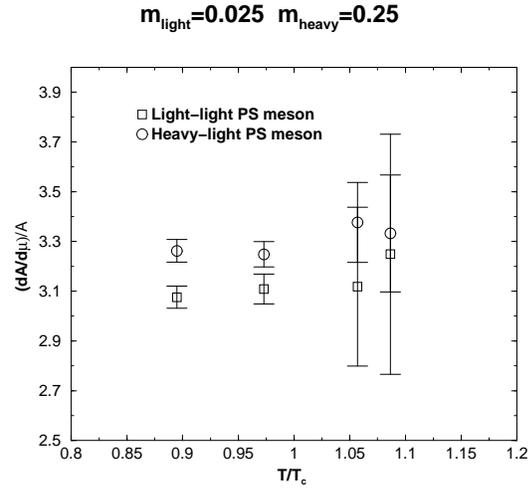,height=6.5cm}}
\vspace{-5mm}
\caption{Response of the coupling A to chemical potential,
$\partial \ln A / \partial \mu$ as a function of $T/T_c$.
}
\end{figure}

Fig. 3 shows the response of the coupling $A$, $\partial \ln A/\partial \mu$ as
a function of $T/T_c$. Both light-light and heavy-light mesons show 
similar values and no apreciable temperature dependence.  

\section{Discussions}
Our preliminary results show remarkable characteristics of the response of
meson masses to chemical potential.
Possible interpretations for  $\partial m/\partial \mu$ of the light-light system are
as follows. The weak response of the mass below $T_c$ indicates a persistence ofthe Nambu-Goldstone
boson nature at least up to $T=0.97T_c$.
Growth of it above $T_c$ is consistent with chiral restoration since the meson
looses the Nambu-Goldstone character.

\vspace{5mm}

Calculations reported here were done on HSP ( NEC ) at INSAM, Hiroshima Univ.  
This work is supported by the Grant-in-Aid for Scientific Research by Monbusho, 
Japan (No.11740159).

\end{document}